\documentclass{ws-procs975x65}

\begin{document}



\title{LORENTZ SYMMETRY VIOLATION IN NEUTRINOS IN CURVED SPACETIME AND ITS CONSEQUENCES
}

\author{BANIBRATA MUKHOPADHYAY
}

\address{Department of Physics,
Indian Institute of Science, 
Bangalore-560012, India 
\email{bm@physics.iisc.ernet.in}}

%


\def\lsim{\lower.5ex\hbox{$\; \buildrel < \over \sim \;$}}
\def\gsim{\lower.5ex\hbox{$\; \buildrel > \over \sim \;$}}

\def\ch{\lower-0.55ex\hbox{--}\kern-0.55em{\lower0.15ex\hbox{$h$}}}
\def\lh{\lower-0.55ex\hbox{--}\kern-0.55em{\lower0.15ex\hbox{$\lambda$}}}

\begin{abstract}
The neutrino propagating in curved spacetimes, e.g. around black holes, in early curved universe, violates 
Lorentz and then CPT symmetry, at least in the local coordinate. This occurs
due to coupling of spin of the neutrino to background spin connection which modifies the underlying 
dispersion relation. This results in the neutrino asymmetry and hence as a consequence
neutrino oscillation probability gets affected.
\end{abstract}

\bodymatter

\section{Introduction}\label{intro}

\enlargethispage*{6pt}

Several astrophysical and cosmological problems are involved with neutrino physics, e.g. 
leptogenesis induced baryogenesis, neutrino cooled accretion disks, $r-$process nucleosynthesis
in supernova explosions etc. A successful description of most of these scenario must 
consider neutrinos in curved spacetime. Naturally, in order to solve such problems, 
one should begin with Dirac equation in curved spacetime.
Indeed in late eighties it was
first pointed out \cite{gas} that the presence of gravitational field
affects neutrino flavors which violates equivalence
principle. Interestingly, the neutrino spin coupled with background spin connection
in curved spacetime gives rise to an interaction which, depending on the nature of spacetime, 
violates Lorentz
and then CPT symmetry\cite{mukh1,mukh2}, atleast in a local coordinate.

\section{Dirac lagrangian in curved spacetime}\label{lag}

Let us recall the fermion Lagrangian density in curved spacetime \cite{schw,mukh1} 
\begin{eqnarray}
{\cal L}=\sqrt{-g}\,\overline{\psi}\left[(i\gamma^a\partial_a-m)+\gamma^a\gamma^5
B_a\right]\psi ={\cal L}_f+{\cal L}_I, ~~~
\label{lagf}
\end{eqnarray}
where
\begin{eqnarray}
B^d=\epsilon^{abcd} e_{b\lambda}\left(\partial_a e^\lambda_c+\Gamma^\lambda_{\alpha\mu}
e^\alpha_c e^\mu_a\right),
\,\,\,\, e^\alpha_a\, e^\beta_b\,\eta^{ab}=g^{\alpha\beta},
\label{bd}
\end{eqnarray}
with the choice of unit $c=\ch=k_B=1$, $m$ be the Majorana mass 
and ${\cal L}_I$ may be a Lorentz violating interaction.
Note that under Lorentz transformation, 
$\overline{\psi}\left(\gamma^0\gamma^5\right)\psi\rightarrow -\overline{\psi}
\left(\gamma^0\gamma^5\right)\psi$: pseudo-scaler, and 
$\overline{\psi}\left(\gamma^i\gamma^5\right)\psi\rightarrow \overline{\psi}
\left(\gamma^i\gamma^5\right)\psi$: pseudo-vector. Hence, if background gravitational
potential is constant, then overall interaction is Lorentz symmetry violating, which 
happens at a suitable curved background\cite{mukh1,mukh2}, splitting
the dispersion energies\cite{mukh2} for neutrino and antineutrino in standard model as
\begin{eqnarray}
E_{\nu} =  \sqrt{({\vec p} - {\vec B})^2 + m^2} + B_0, ~~~~
E_{\overline{\nu}} = \sqrt{({\vec p} + {\vec B})^2 + m^2} -
B_0. \label{edis}
\end{eqnarray}
From eqn. (\ref{lagf}) the effective mass of a neutrino\cite{sm08} 
\begin{eqnarray} 
{\cal M} =\left(\begin{array}{cc}-B_0 & -m \\
-m & B_0\end{array}\right), \,\,{\rm where}\,\,{\cal D}_\mu\equiv(\partial_0,\partial_i+\gamma^5 B_i).
\end{eqnarray}
Therefore, in case of flavor mixing, the mass lagrangian for $\nu_e$ and $\nu_\nu$ mixed by a
Majorana mass $m_{e\mu}$ is\cite{sm08}
\begin{eqnarray}
\nonumber (-g)^{-1/2} {\cal L}_m~&=&~-\frac{1}{2}\left(\nu_{e1}^\dag
m_{e1}\nu_{e1}~+~\nu_{e2}^\dag m_{e2}\nu_{e2}~+~\nu_{\mu1}^\dag
m_{\mu1}\nu_{\mu1}~+~\nu_{\mu2}^\dag m_{\mu2}\nu_{\mu2}\right.\\&+& \left.~\nu_{\mu1}^\dag m_{\mu e} \nu_{e1}~+~\nu_{\mu2}^\dag m_{\mu e} \nu_{e2}~+~\nu_{e1}^\dag m_{\mu e} \nu_{\mu1}~+~\nu_{e2}^\dag m_{\mu e} \nu_{\mu2}\right),\end{eqnarray}
where $m_{e1,2}$, $m_{\mu1,2}$ are gravitationally modified masses of neutrino flavors corresponding
to the electron and muon sectors respectively given by $m_{(e,\mu)1,2}=\mp\sqrt{B_0^2+m_{e,\mu}^2}$.
Therefore, the mass square difference of mass eigenstates, which determines the probability of flavor 
oscillation, is given by\cite{sm08}
\begin{eqnarray}
\Delta M^2=|M_1^2-M_2^2|=|M_3^2-M_4^2| 
=\left(\sqrt{B_0^2+m_\mu^2}+\sqrt{B_0^2+m_e^2}\right) \nonumber
\\
\times\sqrt{\left\{\left(\sqrt{B_0^2+m_\mu^2}-\sqrt{B_0^2+m_e^2}\right)^2+4m_{e\mu}^2\right\}}.
\end{eqnarray}

\section{Consequences}\label{cons}

\subsection{Early universe}
As $E_\nu$ and $E_{\bar\nu}$ are different, there is a possibility of neutrino$-$antineutrino
asymmetry with $\Delta n\sim g_dT^2 B_0$, where $g_d$ is the relativistic degrees of freedom
and $T$ the temperature of
the universe. In early universe with gravy wave perturbation\cite{sm08} 
at $T\sim 10^{15}$GeV, $\Delta n/s\sim 10^{-10}$, what we observe today.
During this phase the flavor oscillation is supposed to take place vigorously,
controlled by gravity completely. 

\subsection{During nucleosynthesis phase}
At the age of universe $t\sim 1$sec, if $B_0\sim 5\times 10^{-2}$eV $\sim m_e, m_\mu, m_{e\mu}$, then
the flavor oscillation probability $P=0.999\sin^2\left(\frac{7\times10^{-4}~{\rm eV^2sec}~t}{4E\ch}\right)$.
For TeV neutrinos gravity effect increases probability two orders of magnitude compared to that in
flat space, while for thermal neutrinos the effect is $1.5$ times.

\subsection{Around primordial black holes}
In this case $B_0$ was already computed\cite{mukh1} as $B_0=4a\sqrt{M}z/{\bar \rho}^2\sqrt{2r^3}$, where
${\bar \rho}^2=2r^2+a^2-x^2-y^2-z^2$, $a$ is the specific angular momentum of the black hole. 
For neutrinos at around $20$ Schwarzschild radii away of the
black hole of mass $M\sim 10^{22}M_\odot$, $B_0\sim 5\times 10^{-2}$eV and oscillation length 
decreases to an order of magnitude compared to that in flat space.

\section{Bounds on $B_a$}

In nonrelativistic limit, gravitational interaction becomes $\vec{s}.\vec{B}$, where $\vec{s}$
be the spin of the fermion. According to the Eot-Wash II experiment\cite{adel}, where 
a macroscopic number of fermions could be polarized in the same direction, the difference in
energy between fermion spins polarized parallel and anti-parallel to $\vec{B}$ is proportional
to $|\vec{B}|$. Therefore, by measuring $E$ experimentally, a bound on $\vec{B}$ can be imposed.

In a parametric material we can measure the net magnetization using a squid\cite{ni},
when external field $|\vec{B}|$ appears as an effective magnetic field $B_{\rm eff}=|\vec{B}|/\mu_B$.
In the experiment, $B_{\rm eff}$ can be probed to be $\sim 10^{-12}$G, which implies $\vec{B}\sim 10^{-23}$GeV.

About $10$ Schwarzschild away from a black hole of mass $mM_\odot$, $B_0\sim 10^{-21}/m$~GeV which
could affect the flavor oscillation length, as obtained in flat space, if $m\sim 10^{-11}$, 
because the mass of a neutrino is $\sim 10^{-10}$GeV. However, a neutrino$-$antineutrino oscillation 
length would be $\gsim 10m$ km in the vicinity of a black hole, when
$B_0$ is very small ($\sim 10^{-22}$GeV for a $10M_\odot$ 
black hole) which is quite compatible with the experimental bound described above. 

For neutrinos in a satellite orbiting Earth with an azimuthal velocity $v_\phi\sim 1$ km/sec,
$B_0\sim 10^{-37}$GeV, is further smaller, which is very hard to detect.

\section{Conclusions}

A neutrino coupling to spacetime curvature violates Lorentz and then CPT symmetry with a
suitable gravitational background. The conditions: (1) spacetime must not be spherical symmetric, 
(2) background curvature coupling is either a constant or an even function of spacetime.
This results in a possible neutrino$-$antineutrino asymmetry and then 
leptogenesis induced baryogenesis based on sphaleron process. This also leads to
the neutrino$-$antineutrino oscillation. Early curved universe is a very feasible situation 
to have this Lorentz symmetry violation occurred. Gravitational field modifies the mass and then the
mass matrix of neutrinos which allows the particle and anti$-$particle to mix. This 
generates neutrino mass varying with time.  Once gravity modifies the mass of neutrinos, 
it might affect flavor oscillation
probability and corresponding oscillation length, depending on the nature of gravity.

\section*{Acknowledgments}
This work is partly supported by a project, Grant No. SR/S2HEP12/2007, funded
by DST, India.





\end{document}